\documentclass[12pt,a4paper,draft]{article}
\usepackage{graphicx}
\usepackage{amsmath,amsbsy,amsfonts}
\usepackage{cite}
\usepackage{appendix}

\newcommand{\A}{\mathcal{A}}
\newcommand{\F}{\mathcal{F}}
\newcommand{\OW}{\mathcal{Q}}
\newcommand{\re}{\mathrm{e}}
\newcommand{\tr}{\mathrm{tr}\,}

\newcommand{\sun}{SU$(N)$}
\newcommand{\pa}{\partial}

\newcommand{\vone}{v_{_{1}}}
\newcommand{\vtwo}{v_{_{2}}}
\newcommand{\vthree}{v_{_{3}}}
\newcommand{\vn}{v_{_{n}}}
\newcommand{\An}{\mathcal{A}_{_{n-1}}}
\newcommand{\Azero}{\mathcal{A}_{_{0}}}
\newcommand{\Aone}{\mathcal{A}_{_{1}}}
\newcommand{\Atwo}{\mathcal{A}_{_{2}}}

\newcommand{\Fu}{\underline{F}}

\newcommand{\fu}{\underline{f}}
\newcommand{\gu}{\underline{g}}
\newcommand{\Bu}{\underline{B}}

\newcommand{\inty}{\int\!{\mathrm{d}^4y}\;}
\newcommand{\intx}{\int\!{\mathrm{d}^4x}\;}
\newcommand{\intz}{\int\!{\mathrm{d}^4z}\;}

\newcommand{\intw}{\int\!{\mathrm{d}^4{\omega}}\;}
\newcommand{\intxy}{\int\!{\mathrm{d}^4x\,\mathrm{d}^4y}\;}
\newcommand{\intyz}{\int\!{\mathrm{d}^4y\,\mathrm{d}^4z}\;}
\newcommand{\intxyz}{\int\!{\mathrm{d}^4x\,\mathrm{d}^4y\,\mathrm{d}^4z}\;}
\newcommand{\intzw}{\int\!{\mathrm{d}^4z\,\mathrm{d}^4\omega}\;}
\newcommand{\intxyw}{\int\!{\mathrm{d}^4x\,\mathrm{d}^4y\,\mathrm{d}^4\omega }\;}
\newcommand{\intxyu}{\int\!{\mathrm{d}^4x\,\mathrm{d}^4y\,\mathrm{d}^4u}\;}

\newcommand{\intxywz}{\int\!{\mathrm{d}^4x\,\mathrm{d}^4y\,\mathrm{d}^4\omega \,\mathrm{d}^4z}\;}

\newcommand{\kzero}{K_{_{0}}}
\newcommand{\kone}{K_{_{1}}}
\newcommand{\ktwo}{K_{_{2}}}

\newcommand{\kn}{K_{_{n}}}

\newcommand{\koxz}{K_{_{0}}(x,z)}

\newcommand{\kozy}{K_{_{0}}(z,y)}
\newcommand{\kozw}{K_{_{0}}(z,\omega)}
\newcommand{\koxw}{K_{_{0}}(x,\omega)}
\newcommand{\kowz}{K_{_{0}}(\omega,z)}

\newcommand{\knone}{K_{_{n-1}}(z,y)}
\newcommand{\kntwo}{K_{_{n-2}}(z,y)}
\newcommand{\kzeroti}{\tilde{K}_{_{0}}(x,y)}
\newcommand{\koneti}{\tilde{K}_{_{1}}(x,y)}
\newcommand{\ktwoti}{\tilde{K}_{_{2}}(x,y)}
\newcommand{\knxz}{\tilde{K}_{_{n-1}}(x,z)}
\newcommand{\kntxz}{\tilde{K}_{_{n-2}}(x,z)}
\newcommand{\knt}{\tilde{K}_{_{n}}(x,y)}

\newcommand{\konexzt}{K_{_{1}}^{^T}(x,z)}

\newcommand{\arhot}{A^{^{T}}_\rho}
\newcommand{\konetab}{K_{_{1}}^{^T{ab}}(x,y)}

\newcommand{\ktwoab}{K_{_{2}}^{ab}(x,y)}
\newcommand{\arhoab}{A^{ab}_\rho(z)}
\newcommand{\kabxy}{K^{ab}(x,y)}

\newcommand{\konet}{K_{_{1}}^{^{T}}}

\newcommand{\ktwottab}{K_{_{2}}^{^{TT}{ab}}(x,y)}

\newcommand{\ooneab}{\mathcal{Q}_{1}^{ab}}

\newcommand{\otwottab}{\mathcal{Q}_{_{2}}^{^{TT}{ab}}(x,y)}
\newcommand{\otwotlab}{\mathcal{Q}_{_{2}}^{^{TL}{ab}}(x,y)}
\newcommand{\otwollab}{\mathcal{Q}_{_{2}}^{^{LL}{ab}}(x,y)}
\newcommand{\yonetab}{\mathcal{Y}_{_{1}}^{{ab}}(x,y)}

\title{The factorisation of glue and mass terms in SU($N$) gauge theories}
\author{Martin~Lavelle, David~McMullan and Poonam~Sharma \\ \\
School of Computing and Mathematics, University of Plymouth \\
Plymouth, PL4 8AA, UK \\
email: poonam.sharma@plymouth.ac.uk \\
}
\date{\today}

\begin{document}
\maketitle

\begin{abstract}

In this paper we investigate the structure of the glue in Zwanziger's gauge invariant expansion for the  $A^2$-type mass term in Yang-Mills theory. We show how to derive this expansion, in terms of the inverse covariant Laplacian, and extend it to higher orders. In particular, we give an explicit expression, for the first time,  for the next to next to leading order term. We further show that the expansion is not unique and give  examples of the resulting ambiguity.
\end{abstract}


\section{Introduction and motivation}
Identifying gauge invariant configurations is an essential first step in exploring the physical consequences of  gauge theories. The standard repertoire of such states is quite limited and usually associated with the appropriate Wilson lines \cite{Wilson:1974sk} or Polyakov loops \cite{Polyakov:1987ez}. However, it is not at all obvious  if these have the best overlap with actual physical configurations \cite{Heinzl:2007cp,Heinzl:2007kx,Heinzl:2008bu,Heinzl:2008tv} of the theory. This motivates us to explore a wider class of  gauge invariant constructions in a pure gauge theory.

The construction of potentially physical mass terms  has generated an interesting debate  about gauge invariant expansions. This has  a long history~\cite{Cornwall:1981zr,Jackiw:1997jga}, even in QED~\cite{Stueckelberg:1957zz}, but has had a resurgence in pure gauge theory over the past decade  \cite{Dudal:2003by,Capri:2005dy,Fischer:2005qe,Greensite:2005yu,Capri:2006ne,Greensite:2006ns,Dudal:2006tp,Gracey:2007ki} which can be traced to a large extent to a seminal paper by Zwanziger~\cite{Zwanziger:1990tn} where the gauge invariance of this construction was addressed. In particular Zwanziger introduced an expansion for a non-abelian mass term as a sum of separately gauge invariant terms. This has been taken up and exploited in several publications, see for example, \cite{Gubarev:2000eu,Gubarev:2000nz,Verschelde:2001ia,Dudal:2002pq,Murakami:2003mw}.

Zwanziger's expansion of the $A^2$ mass term is in terms of powers of the field strengths
(following his notation):
\begin{equation}\label{a0}
\begin{split}
   |A^h|^2&=-\tfrac12\Big(F_{\mu \nu},\Big(\frac{1}{D^2}F_{\mu \nu}\Big)\Big)\\
   &\quad+\Big( \Big(\frac{1}{D^2}F_{\mu \nu}\Big),\Big[\frac{1}{D^2}D_\alpha F_{\alpha\mu},\frac{1}{D^2}D_\beta F_{\beta\nu}\Big]\Big)\\
   &\quad-\Big( \Big(\frac{1}{D^2}F_{\mu \nu}\Big),\Big[\frac{1}{D^2}D_\beta F_{\beta\rho},\frac{1}{D^2}D_\rho F_{\mu\nu}\Big]\Big),
   \end{split}
\end{equation}
where gauge invariance is maintained order by order by the use of the inverse covariant Laplacian. Although Zwanziger introduced this expansion there is no derivation or discussion about how it arises or how is it extended to higher orders or even how unique is the construction. We will offer such a discussion below.

In this paper we will show how physical gluonic configurations can be constructed\cite{Lavelle:1995ty,Bagan:1999jf,Bagan:1999jk}. We will see that there are two possible gauge invariant expressions for the field strength and that the difference between them  will give us insight into Zwanziger's expansion. In particular, this will reveal an abelian gauge structure within non-abelian gauge theory and this will allow for a succinct description of the mass term.

The plan of this paper is as follows. We start in section~2 with a review of the role of dressings in constructing gauge invariant configurations and demonstrate how the abelian substructure arises. This will allow us to write mass terms in QED and QCD in terms of field strengths.   In section 3 the gauge covariant Laplacian and its inverse will be carefully introduced and its main
properties summarised. This will permit us to obtain the first term in Zwanziger's expansion~(\ref{a0}). Then, in section 4, we will clarify how the dressed field can be factorised into the product of two separate gauge invariant structures which will let us  decompose Zwanziger's expansion into the sum of two terms which are separately gauge invariant. In section 5, we give a precise formula for the mass term which then allows us to generate  (\ref{a0}) and then go beyond it. In section 6 we thus obtain the terms which are quadratic and cubic in the field strengths in (\ref{a0}). We also show that there are ambiguities in the expansion which still respect gauge invariance. In section 7,  we calculate the next term in the expansion for the first time, i.e., the one which is quartic in the field strengths. The paper concludes with comments on these results followed by some short appendices to clarify some of the details.

\section{Dressing and the residual abelian gauge structure}
In a pure gauge theory the vector potential $A_\mu^a$ is the basic building block from which all physical configurations are constructed. We will follow the standard
practice of using a matrix notation for this field by introducing anti-hermitian representation matrices $T^a$ for the gauge group and let $A_\mu:=A_\mu^aT^a$. Note, though, that this notation
is representation dependant but for this section we shall take $T^a=\tau^a$ where $\tau^a$ are the appropriate anti-hermitian matrices for the fundamental $N$-dimensional representation
for which $\tr\tau^a\tau^b=-\frac12\delta^{ab}$.\\

The vector potential, though, is not physical as it is not gauge invariant.
Under a gauge transformation described by the group element $U(x)=\re^{g\theta}$ with $\theta(x)=\theta^a(x)\tau^a$,  we have
\begin{equation}\label{a1}
    A_\mu\to A_\mu^U=U^{-1}A_\mu U+\frac1g U^{-1}\pa_\mu U\,.
\end{equation}
The associated field strength $F_{\mu\nu}=\pa_\mu A_\nu-\pa_\nu A_\mu +g[A_\mu,A_\nu]$ then transforms as a field in the adjoint representation of the gauge group
\begin{equation}\label{a2}
    F_{\mu\nu}^U=U^{-1}F_{\mu\nu}U\,.
\end{equation}
To construct gauge invariant gluonic configurations, we first introduce a dressing field $h^{-1}(x)$ out of the gauge fields which transforms under (\ref{a1}) as
\begin{equation}\label{a3}
    h^{-1}\to h^{-1}U \,.
\end{equation}
Using this dressing a gauge invariant gluonic field is given by
\begin{equation}\label{a4}
     A_\mu^h:=h^{-1}A_\mu h+\frac1g h^{-1}\pa_\mu h\,,
\end{equation}
along with a physical field strength
\begin{equation}\label{a5}
    F_{\mu\nu}^h:=h^{-1}F_{\mu\nu}h\,.
\end{equation}
There is a lot of freedom in the choice of the dressing and this reflects the specific physical situation being studied.
For this paper, we shall focus on the dressing that arises from requiring that $\pa^\mu A_\mu^h=0$, and refer to this as the Landau dressing.
 For a detailed discussion of how the dressings are related to gauge fixings we refer the reader to references \cite{Lavelle:1994rh,Ilderton:2007qy}.
 To connect our discussion to those of \cite{Greensite:2005yu,Dudal:2006tp,Gracey:2007ki} we shall take space time to be
  Euclidian so that the Laplacian $\Box=\pa^\mu\pa_\mu$ has a unique Green's function. \\

Solving perturbatively the Landau condition on the dressing results in the solution
\begin{equation}\label{a6}
h^{-1}=\re^v \,,
\end{equation}
where $v=v^a \tau^a$ has an expansion in powers of the coupling as
\begin{equation}\label{a7}
    v=g\vone+g^2\vtwo+g^3\vthree+\cdots\,.
\end{equation}
In this expansion we find that
\begin{equation}\label{a8}
    \vn=\frac1\Box \pa_\mu \An^\mu\,,
\end{equation}
where the first few terms are given by
\begin{equation}\label{a9}
    \Azero^\mu= A^\mu\,,\qquad \Aone^\mu=[\vone,A^\mu]+\tfrac12[\pa^\mu\vone,\vone]
\end{equation}
and
\begin{equation}\label{a10}
\begin{split}
  \Atwo^\mu = \Big([\vtwo,A^\mu]&+\tfrac12[\vone,[\vone,A^\mu]]+\tfrac12[\pa^\mu\vone,\vtwo] \\
    & +\tfrac12[\pa^\mu\vtwo,\vone] -\tfrac16[\vone,[\vone,\pa^\mu\vone]]\Big)\,.
\end{split}
\end{equation}

The resulting dressed potential (\ref{a4}) can then be written in the manifestly transverse form as
\begin{equation}\label{a11}
    A_\mu^h=\left(g_{\mu\nu}-\frac{\pa_\mu\pa_\nu}{\Box}\right)\A^\nu
    \,,
\end{equation}
where we have introduced a generalised potential $\A^\nu$ such that
\begin{equation}\label{a12}
    \A^\nu=A^\nu+g\Aone^\nu+g^2\Atwo^\nu+\cdots
\end{equation}
This potential plays the role in the non-abelian theory of the abelian potential in QED as it allows us to define a gauge invariant transverse field. Its identification reveals an abelian structure within the \sun \ gauge theory since, under the non-abelian gauge transformation (\ref{a1}), the generalised potential $\A^\nu$ transforms as
\begin{equation}\label{a13}
    \A^\nu\to\A^\nu+\pa^\nu\Theta\,,
\end{equation}
where, for example to lowest non-trivial order,
\begin{equation}\label{a1406102011}
    \Theta=\theta+\tfrac12 g[\vone,\theta]+O(g^2)\,.
\end{equation}
We will henceforth refer to $\A^\nu$ as an abelian potential but we stress that this is in a non-abelian Yang Mills theory.\\

Given such an abelian potential it is natural to define an alternative field strength $\F_{\mu\nu}$ as
\begin{equation}\label{a15a}
    \F_{\mu\nu}=\pa_\mu\A_\nu-\pa_\nu\A_\mu \,,
\end{equation}
which is of abelian form but gauge invariant in the non-abelian theory.
However, given the dressing, we have already seen that we can directly define a gauge invariant field strength by (\ref{a5}). To leading order these two field strengths agree but at higher order in the coupling they differ. To understand the relationship between them we note that the directly dressed field strength (\ref{a5}) can be written as the field strength of the dressed potential (\ref{a4}). Hence
\begin{equation}\label{a16a}
     F_{\mu\nu}^h=\pa_\mu A_\nu^h-\pa_\nu A_\mu^h+g[A_\mu^h,A_\nu^h]\,.
\end{equation}
Now using the identity (\ref{a11}) we get the field strength decomposition
\begin{equation}\label{a17a}
    F_{\mu\nu}^h=\F_{\mu\nu}+g[A_\mu^h,A_\nu^h]\,.
\end{equation}

Given this unexpected choice of gauge invariant field strengths in the non-abelian theory, it is important to clarify their role in the description of physical theory. To this end, we note that in QED the dressed field (\ref{a4}) is simply the transverse field $A^{^{T}}_\mu$ where
\begin{equation}\label{a18a}
    A_\mu^{^{T}}=\left(g_{\mu\nu}-\frac{\pa_\mu\pa_\nu}{\Box}\right)A^\nu\,.
\end{equation}
This can then be used to construct in QED a gauge invariant mass-like term:
\begin{equation}\label{a18b}
    \mathcal{M}^2=\intx A_\mu^{^{T}}(x)A_\mu^{^{T}}(x)\,.
\end{equation}
A little algebra shows that this can be written as
\begin{equation}\label{a18c}
    \mathcal{M}^2=-\tfrac12\intx
    F_{\mu\nu}(x)\Big(\frac{1}{\Box}F^{\mu\nu}\Big)(x)\,.
\end{equation}
In much the same way, in a non-abelian theory a similar mass term can be written as
\begin{equation}\label{a18}
   \mathcal{M}^2 =\intx A^{h\,a}_\mu(x) A^{h\,a}_\mu(x)\,,
\end{equation}
and has received much attention in the literature.
From equation (\ref{a11}) we see that we can write the Landau dressed vector potential as
\begin{equation}\label{a19}
    A_\mu^h=\frac{1}{\Box}\pa^\nu \F_{\nu\mu}\,.
\end{equation}
Using this we have from (\ref{a18})
\begin{equation}\label{a20}
 \intx A^{h\,a}_\mu(x) A^{h\,a}_\mu(x)=-\tfrac12\intx \F_{\mu \nu}^a(x)\Big(\frac{1}{\Box}\F^{\mu\nu}\Big)^{\!\!a}\!\!(x)\,.
\end{equation}
Here we see that the generalisation of the QED mass term (\ref{a18c}) to the non-abelian theory is accomplished by replacing the QED field strength $F_{\mu\nu}$  by the physical field strength $\F_{\mu\nu}$.
Having understood the role of the field strength $\F_{\mu\nu}$ in describing the non-abelian mass term, we now turn to a study of the role of the dressed field strength $F_{\mu\nu}^h$. Its role is not immediately obvious, and we will first need to discuss the approach of Zwanziger to the mass term.

The starting point of Zwanziger's analysis is the observation that another possible generalisation of the right hand side of (\ref{a18c}) which maintains gauge invariance is to use the non-abelian field strength $F_{\mu\nu}$ but replace the inverse Laplacian $\frac {1}{\Box}$ with the gauge covariant inverse $\frac{1}{D^2}$. This leads to the first part of Zwanziger's approach to the mass term as given by the first term of the expansion (\ref{a0}). What we will show in the next sections is
 how this is carried out  and that the dressed field strengths $F_{\mu\nu}^h$ play a central role here. This will then allow us to understand the
 relationship between the non-abelian mass term (\ref{a20}) and the gauge covariant Laplacian approach of Zwanziger.

\section{The Gauge Covariant Inverse Laplacian}
Before discussing how the inverse to the gauge covariant Laplacian is defined perturbatively, we first recall how the inverse to the normal Laplacian, $\Box$, is defined. Acting on an element $f(x)$, from a suitable class of test-functions, we define
\begin{equation}\label{b1}
    \frac{1}{\Box}f(x):=\inty \kzero(x,y) f(y)\,,
\end{equation}
where $\kzero(x,y)$ is the Green's function for the Laplacian which in this Euclidian setting we can write as
\begin{equation}\label{b2}
    \kzero(x,y)=-4\pi^2\frac{1}{(x-y)^2}\,.
\end{equation}
The Green's function satisfies $\Box_x\kzero(x,y)=\delta^4(x-y)$
which ensures that (\ref{b1}) is the inverse Laplacian. Note the
subscript $x$ in the Laplacian signifies which variable it acts
upon.

To generalise this we consider the operator obtained by replacing the derivative $\pa_\mu$ with the covariant derivative $D_\mu=\pa_\mu+gA_\mu$. That is, we consider the  gauge covariant Laplacian
\begin{equation}\label{b3}
    D^2:=D^\mu D_\mu= \Box +g(\pa{\cdot}A+2A{\cdot}\pa)+g^2A^2\,.
\end{equation}
To construct the Green's function we need to identify the space of functions that this operator acts on. Given that this is a matrix operator it will act on vectors in the appropriate representation of the gauge group. To signify this we will write the functions as the column vector $\fu$. The cases we will be interested in are when the vectors are in the fundamental, adjoint or tensor products of these representations but for the moment we will not specify the representation.

We now write
\begin{equation}\label{b4}
    \frac{1}{D^2}\fu(x)=\inty K(x,y)\fu(y)\,,
\end{equation}
where $K(x,y)$ is a matrix Green's function and require that
\begin{equation}\label{b5}
    D^2_x K(x,y)=\delta^4(x-y)\,.
\end{equation}
This is solved perturbatively by letting $K(x,y)$ have the expansion
\begin{equation}\label{b6}
    K(x,y)=\kzero(x,y)+g\kone(x,y)+g^2\ktwo(x,y)+\cdots \,.
\end{equation}
We then find that $\kzero(x,y)$ is the free Green's function (\ref{b2}) times the identity matrix as expected while
\begin{equation}\label{b7}
    \Box_x \kone(x,y)+(\pa{\cdot}A+2A{\cdot}\pa_x)\kzero(x,y)=0\,,
\end{equation}
and in general for $n\geq2$
\begin{equation}\label{b8}
    \Box_x {K}_{n}(x,y)+(\pa{\cdot}A+2A{\cdot}\pa_x){K}_{n-1}(x,y)+A^2{K}_{n-2}(x,y)=0\,.
\end{equation}
These equations can be solved in an iterative fashion resulting in
\begin{equation}\label{b9}
    \kone(x,y)=\intz\Big\{\pa^\rho_z\koxz A_\rho(z)\kozy-\koxz A_\rho(z)\pa^\rho_z\kozy\Big\}\,,
\end{equation}
and in general for $n\geq2$
\begin{equation}\label{b10}
\begin{split}
  \kn(x,y)= & \intz\Big\{\pa^\rho_z\koxz A_\rho(z)\knone-\koxz A_\rho(z)\pa^\rho_z\knone \\
    &\qquad \qquad\qquad-\koxz A^2(z)\kntwo\Big\}\,.
\end{split}
\end{equation}
It is useful here to list various key properties of this inverse covariant Laplacian. The derivation of these  are briefly discussed
below  while the full proofs can be found in~\cite{PSthesis}.

The first thing to note is that $1/D^2$ is both a left and right inverse to $D^2$, i.e.,
\begin{equation}\label{b10a}
    D^2\Big(\frac{1}{D^2}\fu\Big)(x)=\Big(\frac{1}{D^2}D^2\fu\Big)(x)=\fu(x)\,.
\end{equation}
This result can be shown in an analogous way to the above construction of the right inverse. That is, we start with a left inverse $1/\tilde{D}^2$ defined by the equation
\begin{equation}\label{b10b}
    \Big(\frac{1}{\tilde{D}^2}D^2\fu\Big)(x):=\inty\tilde{K}(x,y)D^2\fu(y)=\fu(x)\,,
\end{equation}
and then expand in a power series as in (\ref{b6}) to give
\begin{equation}\label{b10c}
      \tilde{K}(x,y)=\kzeroti+g\koneti+g^2\ktwoti+\cdots \,.
\end{equation}
Proceeding in the same way as above we find for $n\geq2$
\begin{equation}\label{bf}
  \begin{split}
  \knt= & \intz\Big\{\pa^\rho_z\knxz A_\rho(z)\kozy-\knxz A_\rho(z)\pa^\rho_z\kozy \\
    &\qquad\qquad\qquad-\kntxz A^2(z)\kozy\Big\}\,.\\
\end{split}
\end{equation}
An inductive proof will then show that order by order we have
$K_{_{n}}(x,y)=\tilde{K}_{_{n}}(x,y)$.\\

A very useful result is that
\begin{equation}\label{b10d}
    \tr\intx \fu(x)\Big(\frac{1}{D^2}\,\gu\Big)(x)=\tr\intx\Big(\frac{1}{D^2}\fu\Big)(x)\,\gu(x)\,.
\end{equation}
For fields in the adjoint representation where $K(x,y)\to
K^{ab}(x,y)$ this identity is equivalent to the result that
\begin{equation}\label{b10e}
    K^{ab}(x,y)=K^{ba}(y,x)\,.
\end{equation}
Again, this identity can be shown to all orders by an induction argument.\\

Note that when we are explicit  about the matrix indices in the Green's function as in $K^{ab}(x,y)$ we mean that we have an expansion
\begin{equation}\label{b10f}
    K^{ab}(x,y)=\delta^{ab}\kzero^{ab}(x,y)+g\kone^{ab}(x,y)+g^2
    \ktwo^{ab}(x,y)+\cdots\,,
\end{equation}
where, for example,
\begin{equation}\label{b10g}
       \kone^{ab}(x,y)=\intz\Big\{\pa^\rho_z\koxz \arhoab\kozy-\koxz \arhoab \pa^\rho_z\kozy\Big\}\,,
\end{equation} and $\arhoab$ is the potential in the adjoint
representation, $\arhoab=-A^\rho_{_{c}} f^{abc}$ where $f^{abc}$ are the structure constants for the group.\\

We now need to clarify the gauge transformation properties of the inverse covariant Laplacian. Under a gauge transformation the covariant Laplacian transforms as
\begin{equation}\label{b12}
    D^2_x\to {U}^{-1}(x)D^2_x \, {U}(x)\,,
\end{equation}
where ${U}(x)$ is the group element in the appropriate
representation that we used to define the covariant derivative.
From the gauge transformation property of the gauge covariant Laplacian (\ref{b12}) and the fact that $K(x,y)$ is the Green's function for both the left and the right inverse of the Laplacian action (\ref{b10a}), (\ref{b10b}) it follows that the associated transformation to the Green's function $K(x,y)$
is then
\begin{equation}\label{b13}
 K(x,y)\to {U}^{-1}(x)K(x,y) {U}(y)\,.
\end{equation}
This means that if we have a field $\Bu$ which transforms in the adjoint representation then so will $\dfrac{1}{D^2} \Bu$, i.e.,
\begin{equation}\label{b13a}
    \Big(\frac{1}{D^2} B\Big)^{\!\!a}\!\!(x)\to (U^{-1})^{ab}(x) \Big(\frac{1}{D^2} B\Big)^{\!\!b}(x)\,.
\end{equation}
While if we have a tensor product of the adjoint representation we get
\begin{equation}\label{b13b}
    \Big(\frac{1}{D^2} B\Big)^{\!\!ab}\!\!(x)\to (U^{-1})^{ab}_{\,\,cd}(x) \Big(\frac{1}{D^2} B\Big)^{\!\!cd}\!\!(x)=(U^{-1})^{ac}(x)(U^{-1})^{bd}(x) \Big(\frac{1}{D^2} B\Big)^{\!\!cd}\!\!(x)\,.
\end{equation}
These properties  allows us to use $K(x,y)$ as a dressing for fields defined at different points $x$ and $y$. In particular if we consider the field strengths $F_{\mu\nu}^a$ at the two points, then working in the adjoint representation we will have the gauge invariant configuration:
\begin{equation}\label{b14}
   \left<\Fu_{\,\mu\nu}(x),K(x,y)\Fu^{\mu\nu}(y)\right>:=F_{\,\mu\nu}^a(x)K^{ab}(x,y)F^{b\,\mu\nu}(y)\,.
\end{equation}
Integrating this expression gives
\begin{equation}\label{b15}
   -\tfrac12\intx F_{\,\mu\nu}^a(x)\Big(\frac{1}{D^2}F^{\mu\nu}\Big)^{\!\!a}\!\!(x):=
  -\tfrac12 \intxy F_{\,\mu\nu}^a(x)K^{ab}(x,y)F^{b\,\mu\nu}(y)\,,
\end{equation}
which is the gauge invariant generalisation of the abelian mass term
(\ref{a18c}) proposed in \cite{Zwanziger:1990tn} by Zwanziger. Indeed, the  term on the left-hand side of (\ref{b15}) is the first term in the expansion of the non-abelian mass operator introduced by Zwanziger and summarised in equation (\ref{a0}).\\

In the next sections we will make clear the relationship between this gauge invariant expression (\ref{b15})  and the non-abelian mass term as described in equation (\ref{a20}). This will allow us to understand how the other terms in the expansion (\ref{a0}) arise and are extended.

\section{Factorising the gauge covariant dressing}
The dressing that arises from the gauge covariant Laplacian is
similar in nature to the dressings used in constructing compact, hadronic
states~\cite{Ilderton:2009jb}. Indeed, if the representation is the
fundamental one, then that is precisely what it is. In general, at
least in the perturbative regime, such dressings factorise into
simple dressings for each individual fermion. For example, in QED where we construct a positronium like state by attaching a
string \cite{Lavelle:1999ki} between  two fermionic fields at points $x$ and $y$, we
have the gauge invariant expression
\begin{equation}\label{c1}
\bar{\psi}(x)\exp\Big(-ie\!\!\int_{y}^x A_{\mu}(z) \
 dz^\mu\Big)\psi(y)\,.
\end{equation}
Although this is clearly gauge invariant, for an arbitrary
contour taken from $x$ to $y$, it is  contour dependent, which would suggest that it  is not fully physical.
Indeed as we will see, it can be viewed as an infinitely excited
positronic state.

To understand the structure of this state better we make a
decomposition of the potential into its transverse and longitudinal
components. That is, adopting  the notation
introduced earlier for the non-abelian theory, we write
$A_\mu=A_\mu^{^{T}}+\pa_\mu \vone$. Substituting this decomposition into (\ref{c1})  we find the state (\ref{c1}) factorises to give
\begin{equation}\label{c116102011}
    \bar{\psi}(x)\exp\big(-ie\vone(x)\big)\exp\Big(-ie\!\!\int_{y}^x A_{\mu}^{^{T}}(z) \
 dz^\mu\Big)\exp\big(ie\vone(y)\big)\psi(y)\,,
\end{equation}
which can  be written in the form
\begin{equation}\label{c2}
   \bar{\psi}(x)h(x)M(x,y)h^{-1}(y)\psi(y)\,,
\end{equation}
where $h^{-1}$ is the abelian dressing constructed out of the longitudinal components of the potential and $M(x,y)$ is a separately gauge invariant and contour dependent contribution to this dressing constructed out of the transverse potential.

In (\ref{c2}) we see how the string type dressing can be factorised
into the product of two separately gauge invariant states with an
additional separately physical but highly excited contribution that
depends on the choice of string. What we want to show now is that
the dressing $K^{ab}(x,y)$ introduced in (\ref{b10f}) has a similar factorisation into the
adjoint dressing needed to compensate for the field strength gauge
transformations and a gauge invariant core. We shall see that this factorisation will lie at the heart of Zwanziger's expansion of the mass term. Before showing this  we
need to first look in more detail at the  dressing (\ref{a5}) used
to make the gauge invariant field strength $F_{\mu\nu}^h$.\\

We have seen from equation (\ref{a5}) how the dressed field strength
$F_{\mu\nu}^h$ is defined directly in the fundamental representation in terms
of the fundamental dressing $h$. It is useful now to see how this is
defined directly in the adjoint representation. To do that we need to look at
$F_{\mu\nu}^h$ in terms of its components, that is,
\begin{equation}\label{c3}
(F_{\mu\nu}^h)^a:=-2 \
\tr(F_{\mu\nu}^h\tau^a)={(h^{-1})}^{ab}F_{\mu\nu}^b\,.
\end{equation}
 In this equation $(h^{-1})^{ab}$ is  the dressing in the
adjoint representation and can be written in terms of the fundamental
dressing as,
\begin{equation}\label{c4}
{(h^{-1})}^{ab}=-2 \ \tr(\tau^a h^{-1}\tau^b h)\,.
\end{equation}
Equivalently, in the adjoint representation we can directly define
\begin{equation}\label{c5}
(h^{-1})^{ab}=(e^{v})^{ab}\,,
\end{equation}
where now $v=v^cT^c$ and $(T^c)_{ab}=-f_{cab}$ are the adjoint representation matrices.
 The adjoint dressing ${(h^{-1})}^{ab}$ in (\ref{c4}) now
transforms under a gauge transformation as
\begin{equation}\label{c6}
{(h^{-1})}^{ab}\to -2 \  \tr(\tau^a h^{-1}U\tau^b U^{-1} h)\,.
\end{equation}
It is easy to check that this becomes the gauge transformation
\begin{equation}\label{c7}
{(h^{-1})}^{ab}\to {(h^{-1})}^{ac}U^{cb}\,,
\end{equation}
where $U^{cb}=-2  \ \tr(\tau^c U\tau^b U^{-1})$ is the adjoint
representation of the transformation.\\

Clearly we can mimic the gauge transformation property (\ref{b13})
of the gauge covariant Laplacian by the factorised dressing
$h^{ac}(x)\kzero(x,y) (h^{-1})^{cb}(y)$. What we now want to
understand is how this factorisation emerges from the full dressing
$K^{ab}(x,y)$. That is, in analogy with the factorisation of the mesonic dressing (\ref{c2}), how is the reduction
\begin{equation}\label{c10}
   -\tfrac12\intx
   F_{\,\mu\nu}^a(x)\Big(\frac{1}{D^2}F^{\mu\nu}\Big)^{\!\!a}\!\!(x)\to-\tfrac12\intx
    F_{\mu\nu}^{h\,a}(x)\Big(\frac{1}{\Box}
    F^{h\,\mu\nu}\Big)^{\!\!a}\!\!(x)\,,
\end{equation}
  achieved in terms of the transverse/longitudinal decomposition of the component fields.\\

To this end we identify
\begin{equation}\label{c11}
    -\tfrac12\intx F_{\mu \nu}^a(x)\Big(\frac{1}{D^2}F^{\mu\nu}\Big)^{\!\!a}\!\!(x)=-\tfrac12\intx F_{\mu\nu}^{h\,a}(x)\Big(\frac{1}{\Box} F^{h\,\mu\nu}\Big)^{\!\!a}\!\!(x)+\mathcal{Q} \,,
\end{equation}
where
\begin{equation}\label{c12}
  \mathcal{Q}=-\tfrac12\intxy F_{\mu \nu}^a(x)\mathcal{Q}^{ab}(x,y){F^{\mu
   \nu\,b}}(y)\,,
\end{equation}
and
\begin{equation}\label{c13}
   \mathcal{Q}^{ab}(x,y)=\kabxy -h^{ac}(x)\kzero(x,y)
   {(h^{-1})}^{cb}(y)\,.
\end{equation}
By construction the operator $\mathcal{Q}$ is gauge invariant as it is the difference of two gauge invariant terms. This means that we must have
\begin{equation}\label{c14}
    \mathcal{Q}^{ab}(x,y) \to
    (U^{-1})^{ac}(x)\mathcal{Q}^{cd}(x,y)(U)^{db}(y)\,,
\end{equation}
under a gauge transformation.\\

Using the perturbative expansions for the various dressings we can write
\begin{equation}\label{c16}
     \mathcal{Q}^{ab}(x,y)=g \mathcal{Q}_{_{1}}^{ab}(x,y)+g^2
     \mathcal{Q}_{_{2}}^{ab}(x,y)+g^3
     \mathcal{Q}_{_{3}}^{ab}(x,y)+\cdots\,,
\end{equation}
which induces an expansion in the operator $\OW$
\begin{equation}\label{c15}
    \mathcal{Q}=
    g\mathcal{Q}_{_{1}}+g^2\mathcal{Q}_{_{2}}+g^3\mathcal{Q}_{_{3}}+\cdots\,,
\end{equation}
although this is not strictly an expansion in the coupling since the field strengths in the definition (\ref{c12}) will also induce powers of the coupling in the $\mathcal{Q}_{_{i}}$ terms.\\

In this expansion it is important to note that it is the sum of all the terms in (\ref{c15}) that
is gauge invariant, individual terms are not. Going beyond this
simple observation lies at the heart of Zwanziger's expansion.\\

To make this point clear let's look at the first and second terms in the expansion (\ref{c16}):
\begin{equation}\label{c18}
\begin{split}
   \mathcal{Q}_{_{1}}^{ab}(x,y)&=\kone^{ab}(x,y) -\big\{ \vone^{ab}(x)-\vone^{ab}(y)\big\}\kzero
   (x,y)\,,\\
   \mathcal{Q}_{_{2}}^{ab}(x,y)&= \ktwoab -\big\{\vone^{ac}(x) \kzero(x,y) \vone^{cb}(y)-\kzero(x,y) \big(\vtwo^{ab}(y)-\vtwo^{ab}(x)\big)\\
 &\qquad\qquad\qquad\qquad\qquad-\kzero(x,y) \big (\tfrac12 \vone^{ac}(x)\vone^{cb}(x)+\tfrac12
 \vone^{ac}(y)\vone^{cb}(y)\big)\big\}\,.
   \end{split}
\end{equation}
The first term $ \mathcal{Q}_{_{1}}^{ab}(x,y)$ can be written as
 \begin{equation}\label{c19}
  \mathcal{Q}_{_{1}}^{ab}(x,y)=\intz\big\{\pa^{\rho}_z\koxz (\arhot)^{ab}(z) \kozy -\koxz (\arhot)^{ab}(z) \pa^{\rho}_z\kozy
    \big\}\,,
\end{equation}
which is simply the transverse part of ${K}_{_{1}}^{ab}(x,y)$, see (\ref{b9}), as it should be for gauge invariance at this order.
In contrast, the next term in the expansion (\ref{c16}) is not purely transverse and not immediately related to the next terms in the expansion of the Laplacian.
\begin{equation}\label{c19a}
     \mathcal{Q}_{_{2}}^{ab}(x,y)\neq {\ktwottab}\,.
\end{equation}
Indeed $\mathcal{Q}_{_{2}}^{ab}(x,y)$ will have a decomposition into transverse-transverse (TT), transverse-longitudinal (TL) and longitudinal-longitudinal (LL) components
\begin{equation}\label{c19b}
     \mathcal{Q}_{_{2}}^{ab}(x,y)=\otwottab+\otwotlab+\otwollab \,,
\end{equation}
which reflects the fact that it is the sum of $g\mathcal{Q}_{_{1}}^{ab}(x,y)+g^2\mathcal{Q}_{_{2}}^{ab}(x,y)$ which now has well defined property described by (\ref{c14}) to this order.\\

It is now clear from (\ref{c19}) that the first term in expansion (\ref{c15})
\begin{equation}\label{c20}
    \mathcal{Q}_{_{1}}=-\tfrac{1}{2}\intxy  F_{\mu \nu}^a(x) \mathcal{Q}_{_{1}}^{ab}(x,y){F^{\mu
   \nu\,b}}(y)\,,
\end{equation}
 is gauge invariant to lowest order in the coupling but at higher order it changes.\\


To make this term in the expansion fully gauge invariant, Zwanziger essentially proposed the replacement
\begin{equation}\label{c21}
\arhot(z)=\dfrac{1}{\Box}\pa_{\beta} (\pa^\beta A^\rho- \pa^\rho
A^\beta)(z) \to \dfrac{1}{D^2}\Big(D_{\beta} F^{\beta\rho}\Big)(z)\,,
 \end{equation}
so that the first term in the expansion of the operator $\OW$ becomes gauge invariant.
At the lowest order in coupling this does not change
$\mathcal{Q}_{_{1}}$ but clearly it  adds new terms to $\OW$ at higher
order which all need to be removed. To understand how this works and hence how  to recover
Zwanziger's next order result, we first need to make precise how the non-abelian mass term (\ref{a20}) is related to Zwanziger's term (\ref{b15}). This will be the topic we turn to next and after that we will see how to implement Zwanziger's resummation of that result to yield a gauge invariant term by term expansion of the mass (\ref{a20}).

%


\section{The Role of the Dressed Field Strength}
Having identified the role of the  field strength $\F_{\mu\nu}$ (\ref{a15a}) in defining a non-abelian mass term (\ref{a20}), and having defined the gauge covariant inverse Laplacian  thus allowing for a perturbative definition of the Zwanziger term (\ref{b15}), we now connect these descriptions by identifying the common role played by the dressed field strength $F_{\mu\nu}^h$ (\ref{a16a}).\\

Already we have seen in (\ref{c11}) that the decomposition of Zwanziger's gauge invariant expression (\ref{b15}) gives a term analogous to the mass term (\ref{a20}) but with the dressed field strength playing the role of the $\F_{\mu\nu}$ field strength. Now using the field strength relation (\ref{a17a}) we shall find a similar decomposition to (\ref{c11}) in the non-abelian mass term (\ref{a20}).

Indeed we see that
\begin{equation}\label{d3}
  -\tfrac12\intx
  \F_{\mu\nu}^a (x) \Big(\frac{1}{\Box}\F^{\mu\nu}\Big)^a(x) =-\tfrac12\intx F_{\mu\nu}^{h\, a}(x)\Big(\frac{1}{\Box} F^{h\,\mu\nu}\Big)^a(x)+\mathcal{P}\,,
\end{equation}
where, using (\ref{a17a}) the gauge invariant term $\mathcal{P}$ is given by
\begin{equation}\label{d4}
\begin{split}
    \mathcal{P}&= \frac{g}{2} \,\intxy ([A_\mu^h,A_\nu^h])^a(x)\, \kzero(x,y)\,(F_{\mu\nu}^h)^a(y) \\
    &\qquad+\frac{g}{2} \,\intxy (F_{\mu\nu}^h)^a(x)\,  \kzero(x,y)\,([A_\mu^h,A_\nu^h])^a(y)\\
    &\qquad\qquad-\frac{g^2}{2}\,\intxy ([A_\mu^h,A_\nu^h])^a(x) \, \kzero(x,y)\,([A_\mu^h,A_\nu^h])^a(y)\,.
    \end{split}
\end{equation}
As before we can introduce a perturbative expansion of this operator:
\begin{equation}\label{d4a}
\mathcal{P}=g\mathcal{P}_{_{1}}+g^2\mathcal{P}_{_{2}}+g^3\mathcal{P}_{_{3}}+\cdots
\end{equation}
However, just as in the corresponding expansion for $\mathcal{Q}$, (\ref{c15}), it is useful to not strictly expand in the coupling but allow in (\ref{d4a}) for the field strength terms $F_{\mu\nu}$ to be kept together. This means, for example, that
\begin{equation}\label{f4}
    \mathcal{P}_{_{1}}=  \,\intxy F_{\mu\nu}^a(x)\,
    \kzero(x,y)\,[A^{^{T}}_\mu,A^{^{T}}_\nu]^{a}(y)\,.
\end{equation}
We again  stress that although $\mathcal{P}$ is gauge invariant, the individual terms like
$\mathcal{P}_{_{1}}$, $\mathcal{P}_{_{2}}$ are not.

Equations (\ref{c11}) and (\ref{d3}) allow us to finally clarify the relation between the mass term (\ref{a20}) and Zwanziger's expression (\ref{b15}). Eliminating the common factor constructed out of the dressed field strength $F^h$ we see that the mass term (\ref{a20}) can alternatively be written as
\begin{equation}\label{d5}
 \mathcal{M}^2 = -\tfrac12\intx F_{\mu \nu}^a(x)\Big(\frac{1}{D^2}F_{\mu
  \nu}\Big)^a(x)+\mathcal{P}-\mathcal{Q}\,.
\end{equation}
where the operators $\mathcal{Q}$ and $\mathcal{P}$ are defined to all orders in perturbation theory by (\ref{c12}) and (\ref{d4}).

This succinct formula is not, though, Zwanziger's proposed expansion of the mass term as the individual terms in the operators $\mathcal{Q}$ and $\mathcal{P}$ are not gauge invariant. In the next section we shall see how to resum these expressions so as to maintain gauge invariance for each term in the new expansions of the operators $\mathcal{Q}$ and $\mathcal{P}$ and this will allow us to generate Zwanziger's expression (\ref{a0}) and then go beyond it.

\section{Recovering Zwanziger's expansion}\label{amb}
The aim now is to investigate how to go from the expansions (\ref{c15}) and (\ref{d4a}) to the resummed expansions
\begin{equation}\label{e1}
\mathcal{Q}=g\mathcal{Y}_{_{1}}+g^2\mathcal{Y}_{_{2}}+g^3\mathcal{Y}_{_{3}}+\cdots
\end{equation}
and
\begin{equation}\label{e2}
\mathcal{P}=g\mathcal{Z}_{_{1}}+g^2\mathcal{Z}_{_{2}}+g^3\mathcal{Z}_{_{3}}+\cdots
\end{equation}
where each term in these expansions will be separately gauge invariant by construction. Note that in these expressions we are not formally expanding in the powers of the coupling and, as we will see, each term is more properly characterised by the power of the field strengths used to construct them. Indeed we expect  $\mathcal{Y}_{_{1}}$ and $\mathcal{Z}_{_{1}}$ to be both cubic in the field strengths, that is, of order $F^3$.

\subsection{$\mathcal{Q}$ to order $F^3$ }
In order to recover Zwanziger's expression for the term $\mathcal{Y}_{_{1}}$ we first need to rewrite (\ref{c19}) in the equivalent form
\begin{equation}\label{e4}
\begin{split}
\mathcal{Q}_{_{1}}^{ab}(x,y)&=-2\intz \pa^{\rho}_x\delta^{ac}\koxz (\arhot)^{cd}(z) \delta^{db}\kozy \,,
  \end{split}
\end{equation}
 where we have simply introduced appropriate colour indices. Then, to impose gauge covariance, we need to make the replacements (\ref{c21}) and
\begin{equation}\label{e4a}
\begin{split}
&\pa^{\rho}_x\delta^{ac}\koxz \to \big(D^{\rho}_x K(x,z)\big)^{ac},\quad \delta^{db}\kozy \to K^{db}(z,y)\,.
\end{split}
\end{equation}
Applying these to (\ref{c20}) we find
\begin{equation}\label{ee5}
\begin{split}
 \mathcal{Y}_{_{1}} &=\intxyz \Big\{F_{\mu \nu}^a(x)\big(D^{\rho}_x
   K(x,z)\big)^{ac}
\Big(\frac{1}{D^2}D_{\beta} F^{\beta\rho}\Big)^{cd}(z)
K^{db}(z,y){F^{\mu \nu}}^b(y)\Big\} \,.
\end{split}
\end{equation}
Using the properties of the inverse Laplacian given by (\ref{b10e}) and integrating (\ref{ee5}) with respect to both $x$ and
$y$ we end up with
\begin{equation}\label{e10}
   \mathcal{Y}_{_{1}} =-\intz \frac {1}{D^2}\Big(D_{\rho}^z F^{\mu \nu}\Big)^c(z)
\Big(\frac{1}{D^2}D_{\beta} F^{\beta\rho}\Big)^{cd}(z) \Big( \frac
{1}{D^2}F^{\mu \nu}\Big)^d(z)\,.
\end{equation}
As shown in Appendix~(\ref{product}) this can be written as
\begin{equation}\label{e12a}
  \mathcal{Y}_{_{1}} =\intz \Big(\frac
{1}{D^2}F^{\mu \nu}\Big)^d(z) \Big[\Big(\frac{1}{D^2}D_{\beta}
F^{\beta\rho}\Big),\Big(\frac {1}{D^2}D_{\rho} F^{\mu
   \nu}\Big)\Big]^d (z)\,,
\end{equation}
which is fully gauge invariant and agrees with the corresponding  term in Zwanziger's expansion (\ref{a0}).

\subsection{$\mathcal{P}$ to order $F^3$ }
Now that we have found  $\mathcal{Y}_{_{1}}$
we return to (\ref{f4}) and follow the same route to find
$\mathcal{Z}_{_{1}}$. In (\ref{f4}) inserting colour indices in an appropriate way, we have
\begin{equation}\label{f1}
    \mathcal{P}_{_{1}}=  \,\intxy F_{\mu\nu}^a(x)\,
    \delta^{ab}\kzero(x,y)\,[A^{^{T}}_\mu,A^{^{T}}_\nu]^{b}(y)\,.
\end{equation}
Now we make the replacements  (\ref{c21}) and
\begin{equation}\label{f1a}
    \delta^{ab}\kzero(x,y) \to K^{ab}(x,y)\,,
\end{equation}
to obtain
\begin{equation}\label{f8}
     \mathcal{Z}_{_{1}}=  \intxy  F_{\mu\nu}^a(x) \kabxy
 \Big[\Big(\frac{1}{D^2} D^\alpha F_{\alpha \mu}\Big),
\Big(\frac{1}{D^2} D^\beta F_{\beta \nu} \Big)\Big]^{b}(y)\,.
\end{equation}
Using (\ref{b10e}) and  (\ref{ap11a}) discussed in the Appendix, we integrate (\ref{f8}) with respect to $x$ to yield:
\begin{equation}\label{f8a}
    \mathcal{Z}_{_{1}}=\inty  \Big(\frac{1}{D^2} F_{\mu\nu}\Big)^b (y)
 \Big[\Big(\frac{1}{D^2} D^\alpha F_{\alpha \mu}\Big),
\Big(\frac{1}{D^2} D^\beta F_{\beta \nu} \Big)\Big]^{b}(y)\,.
\end{equation}

Having obtained  $ \mathcal{Y}_{_{1}}$ (\ref{e12a}) and  $ \mathcal{Z}_{_{1}}$ (\ref{f8a}) we can now write down the expression for the non-abelian mass term to order $F^3$. For this we substitute (\ref{e12a}) and (\ref{f8a}) into (\ref{d5}) to get
\begin{equation}\label{f99}
\begin{split}
\mathcal{M}^2 &= -\tfrac12\intx F_{\mu \nu}^a(x)\Big(\frac{1}{D^2}F_{\mu \nu}\Big)^a(x)\\
  &\qquad +g\intx  \Big(\frac{1}{D^2} F_{\mu\nu}\Big)^a (x)
 \Big[\Big(\frac{1}{D^2} D^\alpha F_{\alpha \mu}\Big),
\Big(\frac{1}{D^2} D^\beta F_{\beta \nu} \Big)\Big]^{a}(x)\\
&\qquad\quad-g\intx \Big(\frac {1}{D^2}F_{\mu \nu}\Big)^a(x)
\Big[\Big(\frac{1}{D^2}D^{\beta} F_{\beta\rho}\Big),\Big(\frac
{1}{D^2}D^{\rho} F_{\mu
   \nu}\Big)\Big]^a (x)+\cdots\,.
  \end{split}
\end{equation}
This is the expected expression for the mass term to order $F^3$ and
is the one obtained by Zwanziger, see equation(\ref{a0}), in \cite{Zwanziger:1990tn}. Each term in the
above expression is fully gauge invariant.

As far as we are aware, although this result has been quoted in many places in the literature, no derivation has been presented.
Note, though, that the expression (\ref{f99}) is not unique. We have made two types of choices that effect the results and choices  then will lead to different expansions while still maintaining gauge invariance. The first type of choice came about from the choice of derivative used in the expression for $\mathcal{Q}_{_{1}}^{ab}(x,y)$ as seen in (\ref{e4}) as compared to (\ref{c19}). It is trivial to see that we could alternatively write
\begin{equation}\label{e4a6102011}
\begin{split}
\mathcal{\tilde{Q}}_{_{1}}^{ab}(x,y)&=-2\intz \delta^{ac}\koxz (\arhot)^{cd}(z) \delta^{db} \pa^{\rho}_z\kozy \,,
  \end{split}
\end{equation}
where we have simply integrated by parts and \ used \ the \ result \ that
$\pa^\rho_x\koxz=-\pa^\rho_z\koxz$. But under the covariant trick  $\pa^\rho_x K(x,z) \neq -\pa^\rho_z K(x,z)$ and hence we get a different gauge covariant expression following this route:
\begin{equation}\label{e12}
  \mathcal{\tilde{Y}}_{_{1}} =\intz \Big(\frac
{1}{D^2}F^{\mu \nu}\Big)^d(z) \Big[\Big(\frac{1}{D^2}D_{\beta}
F^{\beta\rho}\Big),D_{\rho}\Big(\frac {1}{D^2} F^{\mu
   \nu}\Big)\Big]^d (z)\,.
\end{equation}
Another choice arose from the identification (\ref{c21}) of the gauge covariant extension to the transverse field. We could alternatively identify
\begin{equation}\label{c21r}
\arhot(z)=\pa_{\beta} \dfrac{1}{\Box}(\pa^\beta A^\rho- \pa^\rho
A^\beta)(z) \to D_{\beta}\Big( \dfrac{1}{D^2}F^{\beta\rho}\Big)(z)\,,
 \end{equation}
 and this will lead to a new gauge invariant expression in the expansion of the mass term. Indeed following (\ref{c21r}) we would now get
 \begin{equation}\label{e12x}
  \mathcal{\tilde{\tilde{Y}}}_{_{1}} =\intz \Big(\frac
{1}{D^2}F^{\mu \nu}\Big)^d(z) \Big[D_{\beta}\Big(\frac{1}{D^2}
F^{\beta\rho}\Big),\Big(\frac {1}{D^2} D_{\rho}F^{\mu
   \nu}\Big)\Big]^d (z)\,,
\end{equation}
and combining these two choices we would have a term of the form
\begin{equation}\label{e12y}
\mathcal{\tilde{\tilde{\tilde{Y}}}}_{_{1}} =\intz \Big(\frac
{1}{D^2}F^{\mu \nu}\Big)^d(z) \Big[D_{\beta}\Big(\frac{1}{D^2}
F^{\beta\rho}\Big),D_{\rho}\Big(\frac {1}{D^2} F^{\mu
   \nu}\Big)\Big]^d (z)\,.
\end{equation}
Indeed combinations of these would be equally valid.\\

In much the similar way these choices when applied to  $\mathcal{P}_{_{1}}$ (\ref{f1}) lead to the following possibilities:
\begin{equation}\label{f8aa}
    \mathcal{\tilde{Z}}_{_{1}}=\inty  \Big(\frac{1}{D^2} F_{\mu\nu}\Big)^b (y)
 \Big[D^\alpha\Big( \frac{1}{D^2} F_{\alpha \mu}\Big),
\Big(\frac{1}{D^2} D^\beta F_{\beta \nu} \Big)\Big]^{b}(y)\,,
\end{equation}
\begin{equation}
 \mathcal{\tilde{\tilde{Z}}}_{_{1}}=\inty  \Big(\frac{1}{D^2} F_{\mu\nu}\Big)^b (y)
 \Big[\Big( \frac{1}{D^2} D^\alpha F_{\alpha \mu}\Big),
 D^\beta\Big(\frac{1}{D^2} F_{\beta \nu} \Big)\Big]^{b}(y)\,,
 \end{equation}
 and
 \begin{equation}
    \qquad\mathcal{\tilde{\tilde{\tilde{Z}}}}_{_{1}}=\inty  \Big(\frac{1}{D^2} F_{\mu\nu}\Big)^b (y)
 \Big[D^\alpha\Big( \frac{1}{D^2} F_{\alpha \mu}\Big),
D^\beta\Big(\frac{1}{D^2}  F_{\beta \nu} \Big)\Big]^{b}(y)\,.
\end{equation}
 The specific choice used in (\ref{e12a}) and (\ref{f8a}) reflects our aim to derive precisely Zwanziger's form of the expansion. However, other choices might have advantages in specific applications.

In Appendix~\ref{inequality} we show explicitly that, for example,  the expressions (\ref{c21}) and (\ref{c21r}) are not the same and hence the possible expressions listed above are genuinely different gauge invariant expressions for the non-abelian mass term.

\section{Calculation to order $F^{^{4}}$}
We have seen so far how the mass term can be expressed in terms of  powers of the field strengths. Using this we have been able to recover Zwanziger's expression (\ref{a0}) in terms of the quadratic and cubic powers of the field strengths $F^2$, $F^3$ and also exposed the ambiguities in this expression. What we want to do now is, for the first time, construct the next terms in this expansion which will  be quartic in the field strengths.  In our notation from (\ref{d5})--(\ref{e2}) this will correspond to $\mathcal{Z}_{_{2}}-\mathcal{Y}_{_{2}}$.  Just as before we collect these gauge invariant operators once we have identified the transverse residue of the operator at the appropriate order.
To calculate $\mathcal{Y}_{_{2}}$ and $\mathcal{Z}_{_{2}}$ we need to reinstate the higher order modifications we introduced by hand earlier in going from $\mathcal{Q}_{_{1}}\to \mathcal{Y}_{_{1}}$
and $\mathcal{P}_{_{1}}\to \mathcal{Z}_{_{1}}$. This means that we should view this process as the identification of, for example,
\begin{equation}\label{f10}
\begin{split}
    \mathcal{Q}&=g\mathcal{Q}_{_{1}}+g^2\mathcal{Q}_{_{2}}+g^3\mathcal{Q}_{_{3}}+\cdots \\
    &=g\mathcal{Y}_{_{1}}+g\big(\mathcal{Q}_{_{1}}-\mathcal{Y}_{_{1}}\big)+g^2\mathcal{Q}_{_{2}}+\cdots\,,
    \end{split}
\end{equation}
so that now $g^2\mathcal{Y}_{_{2}}$ is the gauge invariant extension of $g\big(\mathcal{Q}_{_{1}}-\mathcal{Y}_{_{1}}\big)+g^2\mathcal{Q}_{_{2}}$. This will only work if this term is only constructed out of the transverse field so that we can use the replacements such as (\ref{c21}).
For the $F^3$ expression (\ref{e4}) this was relatively straightforward as we have seen that $\mathcal{Q}_{_{1}}^{ab}(x,y)$ was purely transverse, but  we have also seen that $\mathcal{Q}_{_{2}}^{ab}(x,y)$  is not as it contains mixed transverse-longitudinal components (\ref{c19b}). What we now need to ensure is that in the $O(g^2)$ parts of the combination $g\big(\mathcal{Q}_{_{1}}-\mathcal{Y}_{_{1}}\big)+g^2\mathcal{Q}_{_{2}}$ only the transverse components survive. That is,
\begin{equation}\label{g1316102011}
    g\big(\ooneab(x,y)-\yonetab\big)^{^{TL}}+g^2\otwotlab =0 \,,
\end{equation}
 and
   \begin{equation}\label{g131a6102011}
    g\big(\ooneab(x,y)-\yonetab\big)^{^{LL}}+g^2\otwollab =0 \,,
\end{equation}
leaving just the contribution from TT components that we will calculate in the next section. This is quite a strong statement of the underlying gauge invariance of this expansion since we have seen that $\mathcal{Y}_{_{1}}$ is ambiguous. What we claim is that the ambiguities exposed in the previous section only contributes to the TT components.\\

In exactly the same way, for the second contribution to the mass term (\ref{d5}), we can write
\begin{equation}\label{f11}
\begin{split}
    \mathcal{P}&=g\mathcal{P}_{_{1}}+g^2\mathcal{P}_{_{2}}+g^3\mathcal{P}_{_{3}}+\cdots \\
    &=g\mathcal{Z}_{_{1}}+g\big(\mathcal{P}_{_{1}}-\mathcal{Z}_{_{1}}\big)+g^2\mathcal{Z}_{_{2}}+\cdots\,,
    \end{split}
\end{equation}
where again $g^2\mathcal{Z}_{_{2}}$ will then be identified with the gauge invariant extension of $g\big(\mathcal{P}_{_{1}}-\mathcal{Z}_{_{1}}\big)+g^2\mathcal{P}_{_{2}}$ assuming there are no residual LL or TL parts.\\

The verification that only transverse fields survive is non-trivial and the full details can be found in \cite{PSthesis}. The essential step in the argument  is the realisation that even in the adjoint expression (\ref{e4}), the replacement (\ref{c21}) involves expression such as
\begin{equation}\label{f11a}
    \Big(\dfrac{1}{D^2}\big(D_{\beta} F^{\beta\rho}\big)\Big)^{ab}(z)=\intw K^{ab}_{\,\,cd}(z,\omega)(D_{\beta} F^{\beta\rho})^{cd}(\omega)\,.
\end{equation}
 In other words the Green's function for the inverse Laplacian now contains fields in the tensor product of the adjoint representation with itself. Dealing with this involves some additional algebraic complexity some of which is illustrated below in the calculation of the TT components.

\subsection{Calculation of  $\mathcal{Y}_{_{2}}$}
Given that there are ambiguities at the $F^3$ level, to make precise how calculations are performed to order $F^4$ we will adopt the expansion chosen by Zwanziger (\ref{a0}) as discussed in the previous section.
To calculate  $\mathcal{Y}_{_{2}}$  which is the gauge invariant extension of $g\big(\ooneab(x,y)-\yonetab\big)^{^{TT}}+g^2\otwottab $ we collect the transverse-transverse (TT) components to obtain
\begin{equation}\label{g13}
\begin{split}
\lefteqn{g\big(\ooneab(x,y)-\yonetab\big)^{^{TT}}+g^2\otwottab }\\
&\qquad =\intz\Big\{2 A_\rho^{^{T}} (x)\koxz \arhot(z) \kozy + 2\pa_\rho^x\konexzt \arhot (z) \kozy \\
&\qquad\qquad\qquad\qquad\qquad-\koxz \arhot(z) \arhot(z) \kozy \Big\}^{ab}\\
&\qquad\qquad-\intzw \Big\{2\pa_\rho^x\koxz \kozw [A_\beta^{^{T}}, \pa_\rho A_\beta^{^{T}}](\omega) \kozy\Big\}^{ab}\,.
\end{split}
\end{equation}
Note that in the above equation the terms like $\konexzt$ need
further expansion as in (\ref{c19}). Sandwiching expression (\ref{g13}) between the two field strengths, as in (\ref{c12}), and performing the substitutions (\ref{e4a}) we obtain:
\begin{equation}\label{g18a}
\begin{split}
\mathcal{Y}_{_{2}}&=-\intx F_{\mu \nu}^a(x)
\Big[\Big(\dfrac{1}{D^2}D_\beta
F^{\beta\rho}\Big),\frac{1}{D^2}\Big[ \Big(\frac{1}{D^2}D_\tau
F^{\tau\rho}\Big),\Big(\frac{1}{D^2}{F^{\mu \nu}} \Big)\Big]\Big]^a(x)
\\
&\quad-2\intx \Big(\frac{1}{D^2}D_\rho F^{\mu\nu}\Big)^a(x)
\Big[\Big(\frac{1}{D^2}D_\sigma
F^{\sigma\lambda}\Big),D_\lambda\frac{1}{D^2}\Big[
\Big(\frac{1}{D^2}D_\alpha
F^{\alpha\rho}\Big),\Big(\frac{1}{D^2}{F^{\mu \nu}}
\Big)\Big]\Big]^a(x)
\\
&\quad+\tfrac12\intx \Big(\frac{1}{D^2} F^{\mu\nu}\Big)^a(x)
\Big[\Big(\frac{1}{D^2}D_\sigma
F^{\sigma\rho}\Big),\Big[ \Big(\frac{1}{D^2}D_\beta
F^{\beta\rho}\Big),\Big(\frac{1}{D^2}{ F^{\mu \nu}}
\Big)\Big]\Big]^a(x)
\\
&\quad-\intx \Big(\frac{1}{D^2}D_\rho
F^{\mu\nu}\Big)^a(x) \Big[\frac{1}{D^2}\Big[\Big(\frac{1}{D^2}D_\alpha
F^{\alpha\beta}\Big),D_\rho\Big(\frac{1}{D^2}D_\tau
F^{\tau\beta}\Big)\Big] ,\Big(\frac{1}{D^2}{F^{\mu \nu}}
\Big)\Big]^a(x)\,,
   \end{split}
\end{equation}
which is now gauge invariant as required.

Some details of the transition from (\ref{g13}) to (\ref{g18a}) in (\ref{g13}) are summarised for one of the terms in Appendix~\ref{derivation}.

\subsection{Calculation of  $ \mathcal{Z}_{_{2}}$}
Using the same strategy as for $\mathcal{Y}_{_{2}} $ we can in a similar way calculate $\mathcal{Z}_{_{2}}$. Using (\ref{f4}) and applying the replacements to next order we obtain
\begin{equation}\label{h1}
\begin{split}
    g\big(\mathcal{P}_{_{1}}-\mathcal{Z}_{_{1}}\big)+g^2\mathcal{P}_{_{2}}&=-\intxy \Big\{F_{\mu\nu}^a(x)\konetab (A_\mu^{^{T}})^{bc}(y)(A_\nu^{^{T}})^{c}(y)\Big\}\\
    &\quad-\intxyw  \Big\{F_{\mu\nu}^a(x)\kzero^{ab}(x,y) \kzero^{bc}(y,\omega)\\
    &\qquad\qquad\qquad\qquad\qquad \times \big([A_\alpha^{^{T}},\pa_\alpha A_\mu^{^{T}}]+[A_\alpha^{^{T}},f_{\alpha\mu}]\big)^{cd}(\omega) (A_\nu^{^{T}})^{d}(y)\Big\}\\
    &-\intxyu \Big\{ F_{\mu\nu}^a(x)\kzero^{ab}(x,y)(A_\mu^{^{T}})^{bc}(y) \kzero^{cd}(y,u)\\
    &\qquad\qquad\qquad\qquad\qquad\qquad \times \big([A_\beta^{^{T}},\pa_\beta A_\nu^{^{T}}]+[A_\beta^{^{T}},f_{\beta\nu}]\big)^{d}(u)\Big\}\\
    &-\intxyw \Big\{ F_{\mu\nu}^a(x)\kzero^{ab}(x,y)\big(\konet(y,\omega) \pa^\alpha f_{\alpha\mu}(\omega)\big)^{bc}(A_\nu^{^{T}})^{c}(y)\Big\}\\
    &-\intxyu \Big\{ F_{\mu\nu}^a(x)\kzero^{ab}(x,y)(A_\mu^{^{T}})^{bc}(y)\big(\konet(y,u) \pa^\beta f_{\beta\nu}(u)\big)^{c}\Big\}\,.
    \end{split}
\end{equation}
After some calculations  we obtain the final result:
\begin{equation}\label{h2a}
    \begin{split}
    \mathcal{Z}_{_{2}}&=\intx \Big(\frac{1}{D^2} F_{\mu\nu}\Big)^{a}(x)\Big[\frac{1}{D^2}\Big[\Big(\frac{1}{D^2}{D_\sigma F^{\sigma \alpha}}\Big),D^\mu\Big(\frac{1}{D^2}{D_\tau F^{\tau \alpha}}\Big)\Big],\Big(\frac{1}{D^2}{D_\rho F^{\rho \nu}}\Big)\Big]^{a}(x)\\
    &+\intx \Big(\frac{1}{D^2} F_{\mu\nu}\Big)^{a}(x)\Big[\Big(\frac{1}{D^2}{D_\sigma F^{\sigma \mu}}\Big),\frac{1}{D^2}\Big[\Big(\frac{1}{D^2}{D_\rho F^{\rho \beta}}\Big),D^\nu\Big(\frac{1}{D^2}{D_\tau F^{\tau \beta}}\Big)\Big]\Big]^{a}(x)\\
    &-2\intx \Big(\frac{1}{D^2}D_\lambda F^{\mu\nu}\Big)^{a}(x)\Big[\Big(\frac{1}{D^2}{D_\alpha F^{\alpha \lambda}}\Big),\frac{1}{D^2}\Big[\Big(\frac{1}{D^2}{D_\sigma F^{\sigma \mu}}\Big),\Big(\frac{1}{D^2}{D_\rho F^{\rho \nu}}\Big)\Big]\Big]^{a}(x)\\
    &-\tfrac{1}{2}\intx \Big(\frac{1}{D^2}\Big[\Big(\frac{1}{D^2}{D_\sigma F^{\sigma \mu}}\Big),\Big(\frac{1}{D^2}{D_\rho F^{\rho \nu}}\Big)\Big]\Big)^{a}(x)\Big[\Big(\frac{1}{D^2}{D_\tau F^{\tau \mu}}\Big),\Big(\frac{1}{D^2}{D_\alpha F^{\alpha \nu}}\Big)\Big]^{a}(x)\,,
    \end{split}
\end{equation}
which is again manifestly gauge invariant.

Hence we see that the non-abelian mass term has an expansion in terms of the gauge invariant combinations of the field strengths,
\begin{equation}\label{i1}
 \mathcal{M}^2 = -\tfrac12\intx F_{\mu \nu}^a(x)\Big(\frac{1}{D^2}F_{\mu
  \nu}\Big)^a(x)+g(\mathcal{Z}_{_{1}}-\mathcal{Y}_{_{1}})+g^2(\mathcal{Z}_{_{2}}-\mathcal{Y}_{_{2}})+\cdots\,,
\end{equation}
where $\mathcal{Z}_{_{1}}-\mathcal{Y}_{_{1}}$ corresponds to Zwanziger's expression (\ref{a0}) and $\mathcal{Z}_{_{2}}-\mathcal{Y}_{_{2}}$ are the new terms given by (\ref{h2a}) and (\ref{g18a}). We stress though that these expansions are not unique and have ambiguities as discussed in Section~\ref{amb}. What we have presented here are the terms closest in form to Zwanziger's expansion  (\ref{a0}) but other ways of representing these gauge invariant quantities exist.

\section{Conclusion}
In this paper we have derived for the first time the next to leading term for Zwanziger's expansion of a gluonic mass expressed in terms of field strengths and the inverse covariant Laplacian. It is quartic in powers of the field strength. This will allow the calculational programme of \cite{Dudal:2006tp} to be extended to higher orders. We have, in this paper, successfully recovered  the  quadratic and cubic terms in Zwanziger's expansion, however, this exposed ambiguities in the construction. Indeed the choice made at the cubic level in the field strength changes the result at quartic level. We have explicitly shown that making different choices leads to different gauge invariant expressions for the mass term.

We started the paper by demonstrating the  role of dressing in constructing the physical gluonic configurations. This revealed  an abelian gauge structure within the non-abelian gauge theory. We saw that there is the possibility to construct a gauge invariant field strength, $\F_{\mu\nu}$, which is of abelian form. A gauge invariant mass term can be constructed in QED in terms of the  field strength. Similarly in QCD we showed that one can use the physical field strengths, $\F_{\mu\nu}$, to construct an analogous mass term. To obtain a gauge invariant term by term expansion for the mass, two ingredients were needed. The first was  the decomposition  of the field strength (\ref{a17a}) which allowed us to write the dressed mass in terms of the field strengths~(\ref{a20}). The  second was the ability to express the terms in Zwanziger's expansion in terms of dressed fields~(\ref{c11}). Mixing these ingredients we were able to give a precise formula for the mass term~(\ref{d5}).

These methods, we feel, can give us insight into the construction of gauge invariant configurations which we hope will be applicable more widely. We can in particular envisage using these ideas to construct new classes of dressings for the hadronic states which may have a good overlap with the ground state~\cite{Heinzl:2008tv}.

\section{Acknowledgements}
DM would like to thank John Gracey for helpful comments. PS would like to thank Plymouth University for a research studentship.

\appendix
\makeatletter
\@addtoreset{equation}{section}
\renewcommand\theequation{\thesection.\@arabic\c@equation}
\makeatother

\section{Useful Lie algebra properties}
In this section of the appendix we highlight some
properties of Lie algebra that are used in the paper.
\subsection{Transformation rules}
It is seen that the Lie algebra valued field strength tensor
$F_{\mu\nu}$ transforms in the adjoint representation as
\begin{equation}\label{ap1}
F_{\mu\nu} \to F_{\mu\nu}^U= U^{-1}F_{\mu\nu}U\,,
\end{equation}
such that in terms of components  $F_{\mu\nu}=F_{\mu\nu}^a \tau^a$
the transformation rule for  $F_{\mu\nu}^a$ is
\begin{equation}\label{ap2}
 F_{\mu\nu}^a\to (U^{-1})^{ab} F_{\mu\nu}^b\,,
\end{equation}
where $(U^{-1})^{ab}=-2 \tr(\tau^a U^{-1} \tau ^b U)$.\\

 In the same way $F_{\mu\nu}^{ab}$ defined by
\begin{equation}\label{ap3}
F_{\mu\nu}^{ab}:=F_{\mu\nu}^{c}(T^c)_{ab}=-F_{\mu\nu}^{c}f_{abc}\,,
\end{equation}
with $f_{cab}=-2\tr ([\tau^c, \tau^a]\tau^b) $, is seen to transform
as
\begin{equation}\label{ap4}
F_{\mu\nu}^{ab} \to (U^{-1})^{ab}_{\,\,cd}(x)
\left(F_{\mu\nu}\right)^{cd}\!\!(x)=(U^{-1})^{ac}(x)(U^{-1})^{bd}(x)
\left(F_{\mu\nu}\right)^{cd}\!\!(x)\,.
\end{equation}
\subsection{Product and commutator in Lie algebra}\label{product}
 Given a matrix-field $A$ in the adjoint representation, $A^{ab}:=-A^c f^{abc}$,
 the action of this matrix field on
some column vector $\Bu$ can be represented as
\begin{equation}\label{ap5}
    (A \Bu)^a:=A^{ab}B^b=-A^c f^{abc} B^b=[A,B]^a\,,
\end{equation}
that is, product in adjoint equals commutator in the Lie algebra.\\

This can be generalised to the tensor product of the adjoint
representation as follows. Defining
\begin{equation}\label{ap6}
 (A)^{ab}_{\,\,cd}=A^e (T^e)^{ab}_{\,\,cd}\,,
\end{equation}
with $(T^e)^{ab}_{\,\,cd}=f^{aec}\delta^{bd}+f^{bed}\delta^{ac}$ we
find
\begin{equation}\label{ap7}
 (A \Bu)^{ab}:=A^{ab}_{\,\,cd} B^{cd}=[A,B]^{ab}=A^{aa'}B^{a'b}-B^{aa'}A^{a'b}\,.
\end{equation}
Both of these results are used explicitly in deriving expressions (\ref{g18a}) and (\ref{h2a}).\\
\subsection{Algebraic identity between the inverse covariant Laplacian and the Green's function}
 In our description we encountered objects such as
\begin{equation}\label{ap11a}
  \Big(\frac{1}{D^2} B\Big)^{a}(x)=\inty K^{ab}(x,y)
 B^{b}(y)\,,
\end{equation}
and
\begin{equation}\label{ap12}
  \Big(\frac{1}{D^2} B\Big)^{ab}(x)=\inty K^{ab}_{\,\,cd}(x,y)
 B^{cd}(y)\,,
\end{equation}
where for such gauge invariant fields $B^{ab}=-f^{abc}B^c$. It is
not immediately obvious that the inverse Covariant Laplacian
preserves the Lie algebra structure displayed in these equations,
that is, in order for
\begin{equation}\label{ap136102011}
    \Big(\frac{1}{D^2} B\Big)^{ab}(x)= -f^{abc}\Big(\frac{1}{D^2}
    B\Big)^{c}(x)\,,
\end{equation}
we require
\begin{equation}\label{a14a06102011}
    f^{ecd}K^{ab}_{\,\,cd}(x,y)=f^{abc}K^{ce}(x,y)\,.
\end{equation}
However this can be shown from the construction of these fields in terms of their components
${K^{(n)}}^{ab}_{\,\,cd}(x,y)$ and ${K^{(n)}}^{ce}(x,y)$ \cite{PSthesis}. Hence we can
set up a dictionary between our construction of the mass term and
Zwanziger's via the identification
\begin{equation}\label{a15}
\Big(\frac{1}{D^2} B\Big)^{ab}(x) C^b(x)=\Big[\Big(\frac{1}{D^2}
B\Big)(x),C(x) \Big]^a\,,
\end{equation}
\begin{equation}\label{a16}
\begin{split}
\Big(\frac{1}{D^2} B\Big)^{ab}_{\,\,cd}(x)
\Big(\frac{1}{D^2}C\Big)^{cd}(x) D^{b}(x)&=\Big[\Big(\frac{1}{D^2}
B\Big)(x),\Big(\frac{1}{D^2}C\Big)(x) \Big]^{ab}D^{b}(x)\\
&=\Big[\Big[\Big(\frac{1}{D^2} B\Big)(x),\Big(\frac{1}{D^2}C\Big)(x)
\Big],D(x)\Big]^{a}\,,
\end{split}
\end{equation}
and
\begin{equation}\label{a17}
    \begin{split}
\Big(\frac{1}{D^2}B\Big)^{ab}(x)\Big(\frac{1}{D^2}C\Big)^{bc}(x)D^{c}(x)&=
\Big(\frac{1}{D^2}B\Big)^{ab}(x)\Big[\Big(\frac{1}{D^2}C\Big)(x),
D(x)\Big]^{b}\\
&=\Big[\Big(\frac{1}{D^2}B\Big)(x),\Big[\Big(\frac{1}{D^2}C\Big)(x),
D(x)\Big]\Big]^{a}\,.
\end{split}
\end{equation}
\section{Discussion of ambiguities}\label{inequality}
Here we show that the  expressions (\ref{c21}) and (\ref{c21r}) are different. We start with (\ref{c21}) and use the perturbative expansion to order $g$ (suppressing colour indices) to get:
\begin{equation}\label{ap13}
\begin{split}
\dfrac{1}{D^2}\Big(D_{\beta} F^{\beta\rho}\Big)(x)&:=\inty K(x,y)(D_{\beta} F^{\beta\rho})(y)\\
&=\inty \Big\{\kzero(x,y) \pa_\beta^y F^{\beta \rho}(y) \\
&\qquad\qquad+g \big(\kone(x,y)  \pa_\beta^y F^{\beta \rho}(y)+\kzero(x,y)  A_\beta(y) F^{\beta\rho}(y)\big)\Big\}\\
&=\inty \Big\{-\pa_\beta^y\kzero(x,y)  F^{\beta \rho}(y) \\
&\qquad\qquad+g \big(-\pa_\beta^y \kone(x,y)  F^{\beta \rho}(y)+\kzero(x,y)  A_\beta(y) F^{\beta\rho}(y)\big)\Big\}\,.\\
\end{split}
\end{equation}
In the similar way (\ref{c21r}) can be expanded perturbatively to order $g$ to yield
\begin{equation}\label{ap14}
\begin{split}
    D_{\beta}\Big( \dfrac{1}{D^2}F^{\beta\rho}\Big)(x)&:= D_\beta^x \inty K(x,y)F^{\beta\rho}(y)\\
    &=\inty \Big\{\pa_\beta^x \kzero(x,y) F^{\beta\rho}(y)\\
    &\qquad\qquad+g \big(  A_\beta(x)\kzero(x,y) F^{\beta\rho}(y)+\pa_\beta^x \kone(x,y)  F^{\beta \rho}(y)\big)\Big\}\,.\\
    \end{split}
\end{equation}
Now we look at the difference between (\ref{ap13}) and (\ref{ap14}) to yield:
\begin{equation}\label{ap15}
    \begin{split}
    \lefteqn{\dfrac{1}{D^2}\Big(D_{\beta} F^{\beta\rho}\Big)(x)- D_{\beta}\Big( \dfrac{1}{D^2}F^{\beta\rho}\Big)(x)}\\
    &\quad =g\inty \kzero(x,y) (A_\beta(y)-A_\beta(x))F^{\beta\rho}(y)\\
   &\quad\quad+g\intyz \Big\{\kzero(x,z)\pa_\lambda^z \kzero(z,y)-\pa_\lambda^z \kzero(x,z)\kzero(z,y)\Big\}\pa_\beta^z A_\lambda(z)F^{\beta\rho}(y)+O(g^2)\,.
    \end{split}
\end{equation}
This is clearly not zero.
\section{Derivation of a sample $F^4$ term}\label{derivation}
 In this section we give the derivation of one of the terms in (\ref{g18a}) and show how it is obtained from the term in (\ref{g13}). To show this we consider the second term of (\ref{g13}) where we use (\ref{e4}) to yield:
\begin{equation}\label{g19}
\begin{split}
    &2\intz\{\pa_\rho^x\konexzt \arhot (z) \kozy\}^{ab} \\
    \qquad\qquad&=-4\intzw \{\pa_\rho^x\koxw
A_\lambda^{^{T}}(\omega)\pa_\lambda^{\omega}\kowz  \arhot (z) \kozy \}^{ab}\,.
\end{split}
\end{equation}
Sandwiching now (\ref{g19}) between the field strengths as in (\ref{c12}) we obtain
\begin{equation}\label{bp1}
2\intxywz F_{\mu \nu}^a(x)  \Big\{ \pa_\rho^x \koxw
A_\lambda^{^{T}}(\omega)\pa_\lambda^{\omega}\kowz \arhot(z) \kozy
\Big\}^{ab} {F^{\mu \nu}}^b(y)\,.
\end{equation}
Expanding the colour indices explicitly  leads to
\begin{equation}
\begin{split}
&2\intxywz \Big\{F_{\mu \nu}^a(x)   (\pa_\rho^x
)^{ac'}K_{_{0}}^{c'c}(x,\omega)
{A^{^{T}cd}_\lambda(\omega)}\\
&\qquad\qquad\qquad\qquad\qquad\qquad\qquad
\times(\pa_\lambda^{\omega}\kowz )^{de}{A^{^{T}ef}_\rho(z)}
K_{_{0}}^{fb}(z,y)
 {F^{\mu \nu}}^b(y)\Big\}\,,
 \end{split}
  \end{equation}
where now the partial derivative $\pa_\rho^x $ acts on the field strength $F_{\mu \nu}^a(x)$ resulting in
 \begin{equation}\label{bpz}
 \begin{split}
 &-2\intxywz \Big\{{K_{_{0}}^{c'c}(x,\omega)}(\pa_\rho^x F_{\mu \nu})^{c'}(x)
{A^{^{T}cd}_\lambda(\omega)}\\
&\qquad\qquad\qquad\qquad\qquad\qquad\qquad
\times(\pa_\lambda^{\omega}\kowz)^{de}{A^{^{T}ef}_\rho(z)}
K_{_{0}}^{fb}(z,y)
 {F^{\mu \nu}}^b(y)\Big\}\,.
 \end{split}
\end{equation}
Using (\ref{b1}) we integrate (\ref{bpz}) with respect to $x$ and $y$ to yield
\begin{equation}\label{bp2}
-2\intzw \Big\{\Big(\frac{1}{\Box}\pa_\rho F_{\mu
\nu}\Big)^{c}(\omega)
{A^{^{T}cd}_\lambda(\omega)}(\pa_\lambda^{\omega}\kowz
)^{de}{A^{^{T}ef}_\rho(z)} \Big(\frac{1}{\Box} {F^{\mu
\nu}}\Big)^f(z)\Big\}\,.
\end{equation}
The last two terms of (\ref{bp2}) become a commutator using the property (\ref{ap5})
\begin{equation}
-2\intzw \Big\{\Big(\frac{1}{\Box}\pa_\rho F_{\mu
\nu}\Big)^{c}(\omega)
{A^{^{T}cd}_\lambda(\omega)}(\pa_\lambda^{\omega}\kowz )^{de}
\Big[{A^{^{T}}_\rho}, \Big(\frac{1}{\Box} {F^{\mu
\nu}}\Big)\Big]^e(z)\Big\}\,,
\end{equation}
where now integrating the above equation with respect to $z$ gives
\begin{equation}\label{bp22}
-2\intw \Big\{\Big(\frac{1}{\Box}\pa_\rho F_{\mu
\nu}\Big)^{c}(\omega)
{A^{^{T}cd}_\lambda(\omega)}\Big(\pa_\lambda^{\omega}\frac{1}{\Box}
\Big[{A^{^{T}}_\rho}, \frac{1}{\Box} {F^{\mu
\nu}}\Big]\Big)^d(\omega)\Big\}\,.
\end{equation}
Again using the property (\ref{ap5}) in (\ref{bp22}) results in
\begin{equation}
-2\intw \Big\{\Big(\frac{1}{\Box}\pa_\rho F_{\mu
\nu}\Big)^{c}(\omega)
\Big[{A^{^{T}}_\lambda},\pa_\lambda^{\omega}\frac{1}{\Box}
\Big[{A^{^{T}}_\rho}, \frac{1}{\Box} {F^{\mu
\nu}}\Big]\Big]^c(\omega)\Big\}\,.
\end{equation}
Now we make the replacement (\ref{c21}) and $\frac{1}{\Box}\to \frac{1}{D^2}$  at this order to obtain
\begin{equation}\label{bp3}
   -2\intw \Big(\frac{1}{D^2}D_\rho F^{\mu\nu}\Big)^c(\omega)
\Big[\Big(\frac{1}{D^2}D_\sigma
F^{\sigma\lambda}\Big),D_\lambda\frac{1}{D^2}\Big[
\Big(\frac{1}{D^2}D_\alpha
F^{\alpha\rho}\Big),\Big(\frac{1}{D^2}{F^{\mu \nu}}
\Big)\Big]\Big]^c(\omega)\,,
\end{equation}
which is the required expression and corresponds to the second term of
(\ref{g18a}).\\
Note that in deriving the above result we have made use of the choices  (\ref{c21}) and (\ref{e4}) but we could equally well have used the other choices which would then lead to different result.


\end{document}